\documentclass[aps,prl,twocolumn,amsfonts,showpacs,superscriptaddress]{revtex4-1}
\usepackage{verbatim,epsfig,amsmath,amssymb,bm,epsf,graphicx,psfrag,bbold,amsthm,amsfonts}
\usepackage{hyperref}
\usepackage{enumitem}
\usepackage{framed}
\setlist{nosep}
\usepackage[all]{xy}
\usepackage{color}
\usepackage[utf8]{inputenc}
\usepackage{natbib}

\usepackage{tikz-cd}
\usepackage{verbatim}
\usepackage{leftidx}

\newcommand{\abs}[1]{\left|#1\right|}

\newcommand{\p}{\partial}

\newcommand{\coker}{\operatorname{coker}}

\newcommand\bT {{\mathbb T}}

\newcommand\bI {{\mathbb I}}

\newcommand\bN {{\mathbb N}}

\newcommand\bR {{\mathbb R}}

\newcommand\bS {{\mathbb S}}
\newcommand\RP {{\mathbb
{RP}}}
\newcommand\bZ {{\mathbb Z}}

\newcommand\beq {\begin{equation}}
\newcommand\eeq {\end{equation}}
\newcommand\beqa {\begin{equation}\begin{array}}
\newcommand\eeqa {\end{array}\end{equation}}
\newcommand\bal {\begin{align}}
\newcommand\eal {\end{align}}
\newcommand{\bea}{\begin{eqnarray}}
\newcommand{\eea}{\end{eqnarray}}

\newcommand{\innerproduct}[2]{\langle #1| #2\rangle}

\newcommand{\ztwo}{\mathbb{Z}_2}

\newcommand{\basesp}{M}
\newcommand{\targetsp}{X}

\theoremstyle{plain}
\newtheorem{thm}{Theorem}
\newtheorem{prop}[thm]{Proposition}

\theoremstyle{definition}

\theoremstyle{remark}

\begin{document}

\title{On the classification of topological defects and textures}

\author{J.~P.~Ang}
\email{jianpeng.ang@gmail.com (Corresponding author)}
\affiliation{Department of Physics and Astronomy, State University of New York at
	Stony Brook, Stony Brook, NY 11794-3840, USA}
\affiliation{C. N. Yang Institute for Theoretical Physics, Stony Brook, NY 11794-3840, USA} 
\author{Abhishodh Prakash}
\email{abhishodh.prakash@icts.res.in}
\affiliation{Department of Physics and Astronomy, State University of New York at
	Stony Brook, Stony Brook, NY 11794-3840, USA}
\affiliation{C. N. Yang Institute for Theoretical Physics, Stony Brook, NY 11794-3840, USA} 
\affiliation{International Centre for Theoretical Sciences (ICTS-TIFR),
Tata Institute of Fundamental Research,
Shivakote, Hesaraghatta,
Bangalore 560089, INDIA}
\date{\today}
\preprint{YITP-SB-18-30}

\begin{abstract}
Ordered phases resulting from spontaneously broken continuous symmetries are effectively described by sigma models of maps to the coset space of Goldstone modes. A classic problem is to classify the topological sectors of the sigma model. In simple cases, this can be expressed in terms of homotopy groups of the target space; but in general, it is a complicated affair -- even in two dimensions -- and a general method is lacking. In this letter, we introduce a technique to systematically classify topological sectors of sigma models in various dimensions using a framework based on higher categorical generalizations of the fundamental group. As a demonstration, we recover some known results and obtain new ones. The technique and relevant mathematical structures are described only qualitatively in this letter; details can be found in our companion paper~\cite{APandJP}.
\end{abstract}

\maketitle
 
\section{Introduction and summary}
A sigma model describes the physics of scalars taking values in a target manifold. In two dimensions, it describes the motion of strings in a background geometry, and is conformal when the background geometry solves the field equations of general relativity. In dimensions greater than two, the sigma model is superficially non-renormalizable and thus serves as an effective description of low energy physics. For example, sigma models describe the spontaneously broken symmetric phases of \emph{ordered media}~\cite{Mermin1979_RevModPhys.51.591_Defects}, such as magnets, superfluids and liquid crystals. When a continuous global symmetry group $G$ of the system is spontaneously broken to a subgroup $H$, the target space of the sigma model is the \emph{coset} space $G/H$ of \emph{Goldstone modes}~\cite{Nambu1960_PhysRev.117.648_goldstone,Goldstone1961,GoldstoneSalamWeinberg1962_PhysRev.127.965_goldstone}, i.e. \emph{order-parameter} space.

In this letter, we address the problem of describing the topological sectors of the sigma model, i.e. the different field configurations which cannot be continuously deformed into one another. Mathematically, these are the homotopy classes, denoted $[\basesp,\targetsp]$, of maps from a spacetime $\basesp$ to a target space $\targetsp$. In the context of ordered media, topological sectors are known as textures or defects. Skyrmions~\cite{SKYRME1962556} are particular examples of non-trivial textures in sigma models. Originally intended as a model of the nucleon, skyrmions were later observed in various physical systems such as Bose-Einstein condensates~\cite{Khwaja2001skyrmions} and liquid crystals~\cite{fukuda2011skyrmions}. Meanwhile, a defect is a lower dimensional subspace $\Sigma$ on which the field may be singular. This means that the effective field is insufficient to describe the dynamics on $\Sigma$ and should be replaced by a microscopic description. Defect sectors are distinct homotopy classes $\left[\basesp, \targetsp\right]$ of field configurations, but $\basesp$ in this case is the complement of the defect $\Sigma$ in spacetime. Vortices are examples of defects and have been observed in superfluids~\cite{experiment_vortex_superfluid}, superconductors~\cite{experiment_vortex_superconductor} and Bose-Einstein condensates~\cite{experiment_vortex_BEC}. For clarity of exposition, we focus on the example of ordered media, where $\targetsp=G/H$, but our results will be general. We briefly mention that topological sectors of gauge theories, thought of as sigma models to the classifying space $BG$ of the gauge group $G$, correspond to the different principal $G$-bundles over $\basesp$; and the topological sectors in string theory correspond to configurations where the string worldsheet wraps non-contractible cycles in the target space.

We assume that both $\basesp$ and $\targetsp$ are connected. In this situation, we can consider a restricted version of this problem, considering based maps, i.e. maps sending a fixed base point on $\basesp$ to some fixed point in $\targetsp$, and based homotopies, i.e. homotopies anchored on the base point. The resulting based homotopy classes will be denoted $\left[\basesp,\targetsp\right]_0$. Relaxing the based condition corresponds to taking a quotient by $\pi_1(\targetsp)$, allowing us to recover $\left[\basesp,\targetsp\right]$ from $\left[\basesp,\targetsp\right]_0$.

In several cases, $\basesp$ may be $d$-dimensional but be homotopy equivalent to a lower dimensional space, such as when it can be continuously shrunk to a lower dimensional \emph{deformation retract}. The smallest dimension amongst all spaces homotopy equivalent to $\basesp$ is called the \emph{homotopy dimension} of $\basesp$. For example, the result eq.(\ref{eq:circle textures}) for textures on a circle also applies to the classification of point defects on $\mathbb{R}^2$ and of line defects on $\mathbb{R}^3$. Similarly, the classification of point defects on $\mathbb{R}^3$ can be reduced to that of textures on $\bS^2$. For that reason, it is more natural to consider the codimension rather than dimension of defects -- codimension $n+1$ defects are equivalent to textures on a linking sphere $\bS^n$, which are classified by $\pi_n(\targetsp)$. 

When $\basesp$ has boundaries, the choice of boundary conditions for the field can sometimes be studied by replacing the original $\basesp$ with an appropriate compact one. For instance, textures on $\bR^d$ which asymptote to a given constant at infinity can be thought of as textures on $\bS^d$, and textures on the unit cube $\bI^d$ with periodic boundary conditions can be thought of as textures on the torus $\bT^d$.

Let us motivate our approach using the example of textures on a circle $\basesp=\bS^1$. The classes $\left[\bS^1,\targetsp\right]_0$ correspond to based maps of $\bS^1$ into $\targetsp$, which is the fundamental group $\pi_1 (\targetsp)$. Thus, based maps from a circle are classified tautologically by $\pi_1(\targetsp)$. This result can be recovered in a different way as
\begin{equation}~\label{eq:1dconnected}
    \left[\bS^1,\targetsp\right]_0 \cong \left[\pi_1(\bS^1),\pi_1(\targetsp)\right],
\end{equation}
where $[G,H]$ denotes the group homomorphisms from $G$ to $H$. Since $\pi_1(\bS^1)\cong\mathbb{Z}$ and $[\mathbb{Z},H]$ is completely determined by the choice of element in $H$ that is assigned to the generator of $\mathbb{Z}$, we recover $\left[\bS^1,\targetsp\right]_0 \cong \pi_1(\targetsp)$. Relaxing the based condition on eq.(\ref{eq:1dconnected}) amounts to identifying elements of $\pi_1(\targetsp)$ up to inner automorphism~\cite{Mermin1979_RevModPhys.51.591_Defects}
\begin{equation}\label{eq:circle textures}
    \left[\bS^1,\targetsp\right]  \cong  \pi_1(\targetsp)/\operatorname{Inn}(\pi_1(\targetsp)).
\end{equation}
This is the well known result~\cite{Mermin1979_RevModPhys.51.591_Defects} that textures on a circle are classified by conjugacy classes of $\pi_1(\targetsp)$. More generally, a space with two codimension 2 defects retracts onto the figure of eight, or the wedge sum\footnote{The wedge sum $X\vee Y$ of two spaces $X$ and $Y$ is the union of the spaces identified at one point, $X\cup Y/x\sim y$, where $x\in X$ and $y\in Y$.} of two circles. A similar analysis shows that the topological sectors are classified by two conjugacy classes $\pi_1(\targetsp)/\operatorname{Inn}\pi_1(\targetsp) \times \pi_1(\targetsp)/\operatorname{Inn}\pi_1(\targetsp)$.

In the latter point of view eq.(\ref{eq:1dconnected}), we reinterpret the topological problem of homotopy classification as the combinatorial problem of counting homomorphisms between algebraic objects. We would like to generalize this point of view to a higher dimensional spacetime $\basesp$. The fundamental group $\pi_1$ captures all the homotopy information of a $1$-dimensional space; in dimension $d$ we should generalize that to some algebraic object which captures all homotopy type information of a $d$-dimensional space. A suitable object is the fundamental crossed $d$-cube, denoted $\Pi_{\le d}$, which is a higher categorical generalization of the fundamental group~\cite{ellis1988}. The definition of crossed cubes is technical and the details can be found in our companion paper~\cite{APandJP}. The main result from the study of crossed cubes which we shall use is the following~\cite{Ellis1993}.
\begin{framed}
\begin{prop}  \label{prop:main}
Topological sectors of a sigma model with $d$-dimensional spacetime $\basesp$, for $d=2,3$, to some target space $\targetsp$ are in bijection with homotopy classes of homomorphisms between \emph{fundamental crossed $(d-1)$-cubes} $\Pi_{\le d}$ of $\basesp$ and $\targetsp$:
 \begin{align} \label{eq:d dim based sectors}
     \left[\basesp,\targetsp\right]_0 &\cong \left[\Pi_{\le d} (\basesp), \Pi_{\le d} (\targetsp)\right]_0 \\ \label{eq:d dim based sectors 2}
     \left[\basesp,\targetsp\right] &\cong \left[\Pi_{\le d} (\basesp), \Pi_{\le d} (\targetsp)\right]
 \end{align}
 \end{prop}
\end{framed}
On the right hand sides of equations (\ref{eq:d dim based sectors}) and (\ref{eq:d dim based sectors 2}), $\left[\Pi_{\le d} (\basesp), \Pi_{\le d} (\targetsp)\right]_0$ (resp. $\left[\Pi_{\le d} (\basesp), \Pi_{\le d} (\targetsp)\right]$) denotes the homomorphisms between the crossed $(d-1)$-cubes modulo a notion of based equivalence (resp. free equivalence) between them. The based and free homotopy classes are related by $\left[\Pi_{\le d}(\basesp), \Pi_{\le d} (\targetsp)\right]\equiv \left[\Pi_{\le d} (\basesp), \Pi_{\le d} (\targetsp)\right]_0/\pi_1(\targetsp)$, just as in the $1$-dimensional case.

Note that proposition \ref{prop:main} is essentially a generalization of equations (\ref{eq:1dconnected}) and (\ref{eq:circle textures}), with the exception that the last step of taking the quotient by equivalences between homomorphisms was not necessary in the $1$-dimensional case. Every map $\phi:\basesp\to\targetsp$ induces a homomorphism $\Pi_{\leq d}(\basesp)\to\Pi_{\leq d}(\targetsp)$ of fundamental crossed $(d-1)$-cubes, with homotopic maps inducing equivalent homomorphisms and non-homotopic maps inducing non-equivalent homomorphisms. Proposition \ref{prop:main} says that the converse is also true for $d=2,3$: each homomorphism of crossed $(d-1)$-cubes is indeed realized by some map $\basesp\to\targetsp$. In one dimension, it is possible to physically interpret a homomorphism of fundamental groups as the cycle along which the circle wraps; similarly, it will be possible to interpret a homomorphism of fundamental crossed $d$-cubes as the cycles along which the cells of $\basesp$ wrap. Hence, it will be possible to physically interpret the textures obtained in this way.

To keep this letter concise and accessible to a large audience, we do not provide definitions for all the terms, nor provide proofs. Instead, we motivate proposition \ref{prop:main} and apply it to several examples. Proofs and mathematical details are substantial and are presented in the companion paper~\cite{APandJP}, to which we direct the interested reader.

The rest of the paper is organized as follows. First, we consider $\basesp$ of homotopy dimension 2 where the classification eq.(\ref{eq:d dim based sectors}) can be interpreted in terms of cohomology (see proposition \ref{prop:2d}). We then study textures on compact two dimensional manifolds as well as loop and knot defects of Heisenberg magnets and nematic liquid crystals. Second, we consider $\basesp$ of homotopy dimension 3 where a cohomological interpretation is possible only when $\pi_2(\targetsp)=0$ (see proposition \ref{prop:Whitehead}). We study textures of ordered media both satisfying and not satisfying this constraint. Finally, we comment about how our work relates to other instances of the use of higher groups in physics.
 
\section{$\basesp$ of homotopy dimension 2}
\subsection{Definitions}
\begin{figure*}
    \centering
    \begin{tabular}{ccc}
    \raisebox{-0.5\height}{\includegraphics[width=0.15\textwidth]{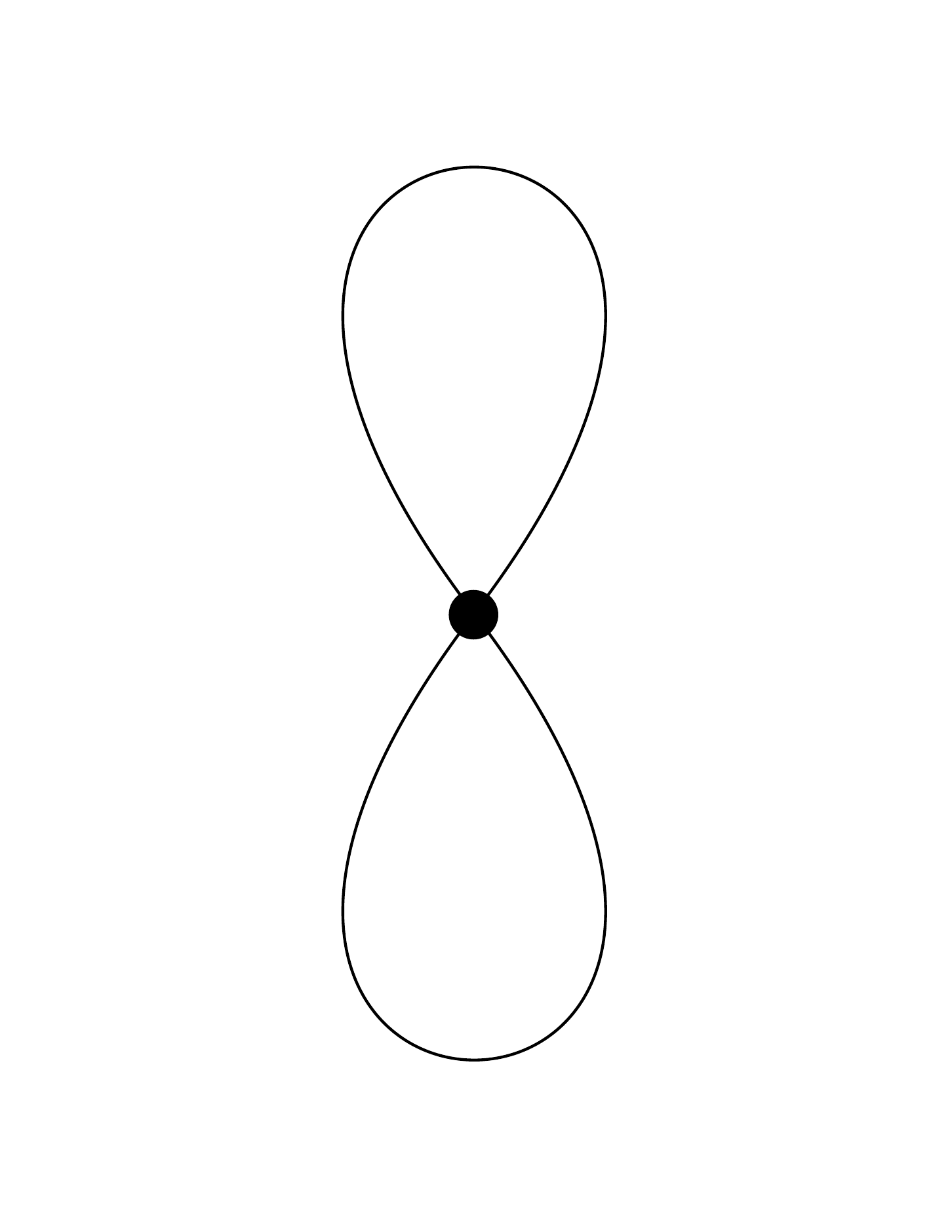}}~~~~~ &
    \raisebox{-0.5\height}{\includegraphics[width=0.2\textwidth]{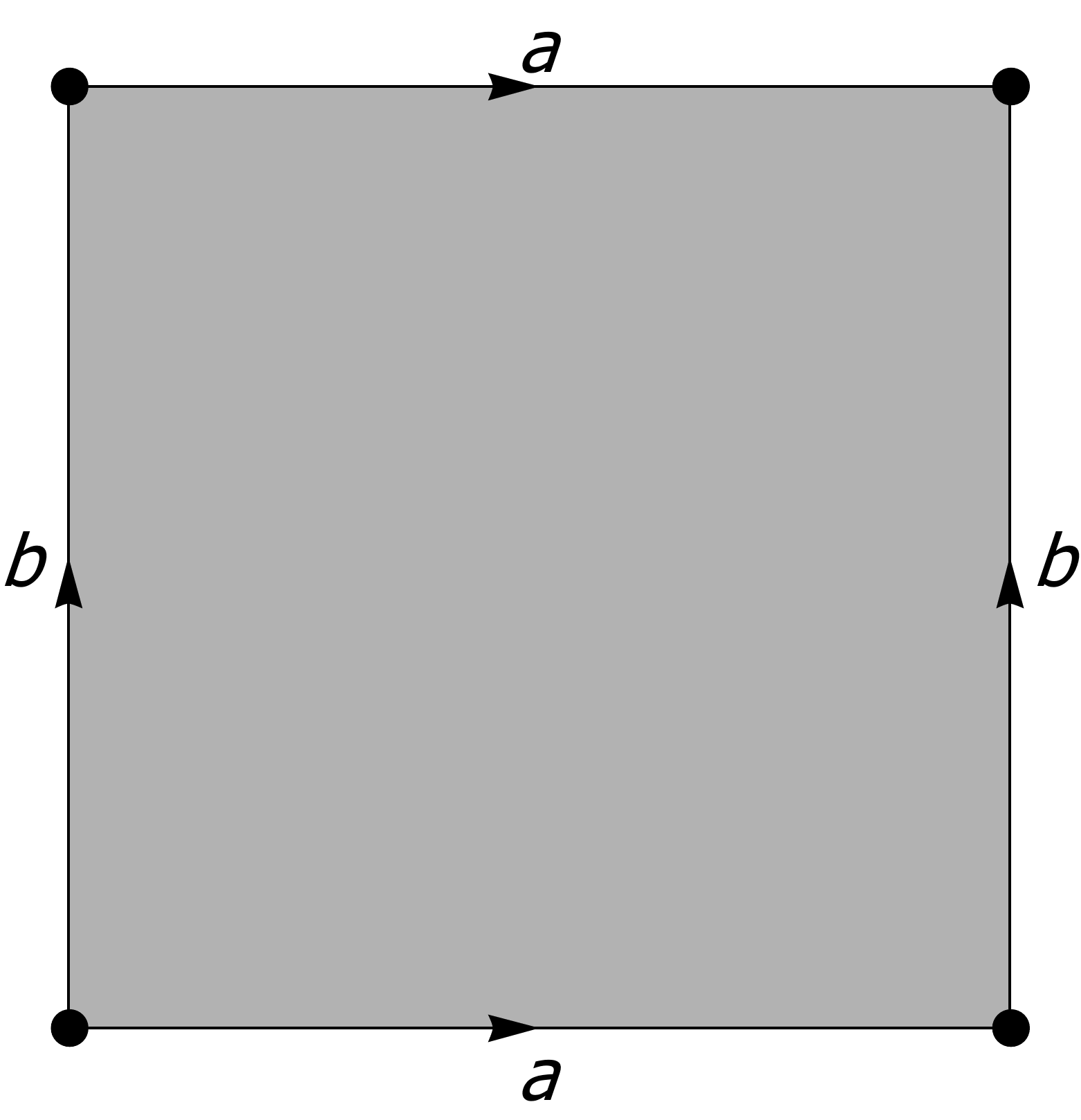}} &
    \raisebox{-0.5\height}{\includegraphics[width=0.4\textwidth]{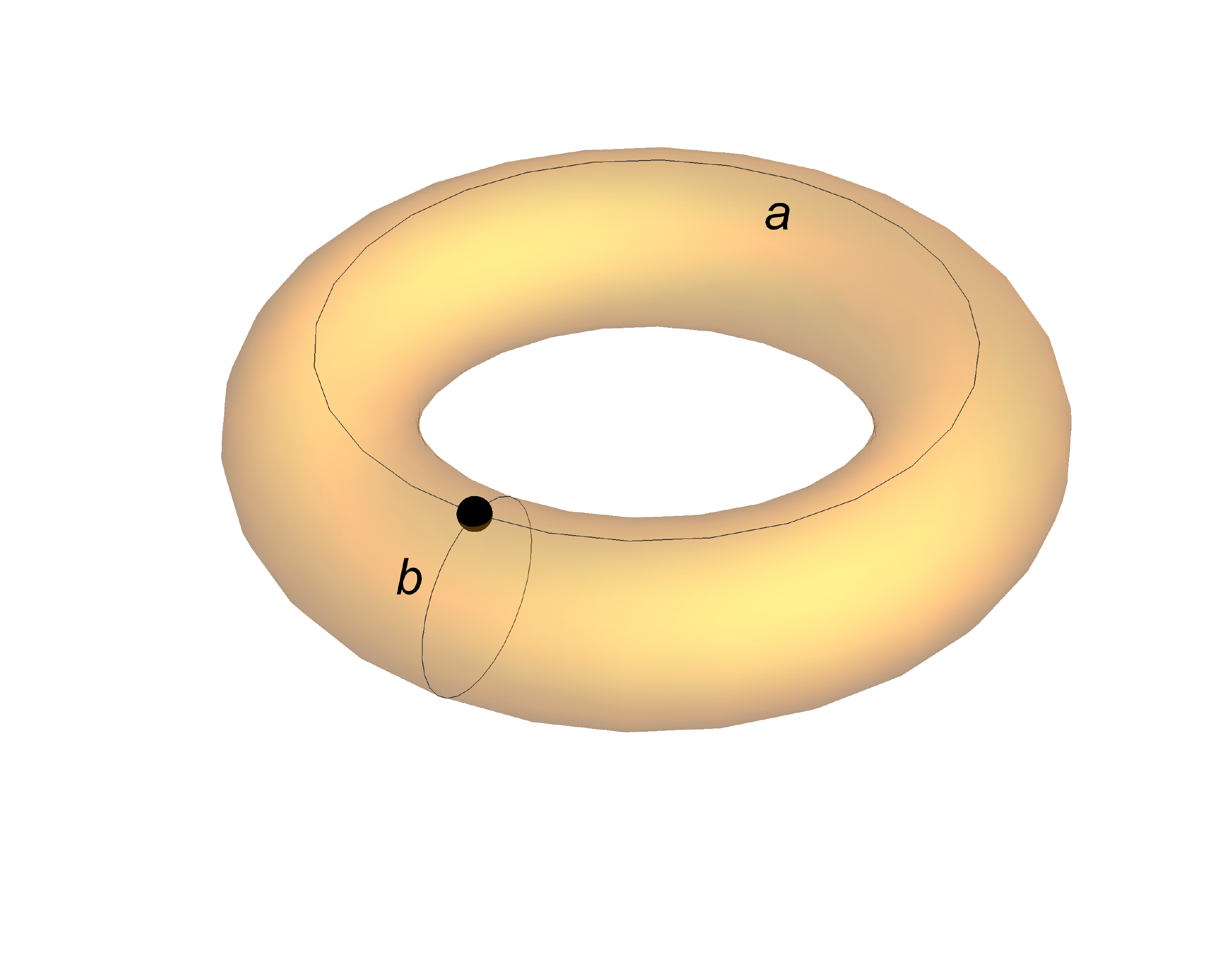}}
    \end{tabular}
    \caption{Left: a figure of eight formed by gluing two 1-cells $a,b$ to one 0-cell. Middle, right: a torus formed by gluing one 2-cell to the figure of eight via $aba^{-1}b^{-1}$. Thus, the 1-skeleton of the torus $\bT^2$ is the figure of eight $\bS^1\vee\bS^1$.}
    \label{fig:cwcomplex}
\end{figure*}

The algebraic object that captures the homotopy information of a two dimensional space is the \emph{fundamental crossed module}, or fundamental crossed $1$-cube, $\Pi_{\le 2}(\basesp)$. This depends on a CW structure on $\basesp$; roughly speaking it is a prescription of constructing $\basesp$ by gluing cells of increasing dimension. The subspace created by gluing all the cells up to those of dimension $n$ is known as the $n$-skeleton, denoted $\basesp^n$. See figure \ref{fig:cwcomplex} for an example, and see e.g.~\cite{hatcher2003} for an introduction to CW complexes.

The fundamental crossed module $\Pi_{\leq 2}(\basesp)$ of $\basesp$ is the boundary map
\beq \p:\pi_2(\basesp,\basesp^1) \to \pi_1(\basesp^1) \eeq
from the second homotopy group of $\basesp$ relative to its 1-skeleton $\basesp^1$ to the fundamental group of $\basesp^1$. When $\basesp$ is two dimensional, $\Pi_{\leq2}(\basesp)$ describes $\basesp$ up to homotopy~\cite{whitehead1949b}. Hence, all of its homotopy data can be extracted from $\Pi_{\leq2}(\basesp)$; in particular the first two homotopy groups
\bea \pi_1(\basesp) &=& \coker\p = \pi_1(\basesp^1)/\p\pi_2(\basesp,\basesp^1), \\
\pi_2(\basesp) &=& \ker\p. \eea
The fundamental crossed module contains more information than just the homotopy groups; this is discussed in detail in~\cite{APandJP}. A systematic prescription for computing the fundamental crossed module is also presented there.

To compute eq.(\ref{eq:d dim based sectors}) and (\ref{eq:d dim based sectors 2}) we have to count crossed module homomorphisms
\beq \begin{tikzcd}
\pi_2(\basesp,\basesp^1) \ar[r,"\p_\basesp"] \ar[d,"\phi_2"]& \pi_1(\basesp^1) \ar[d,"\phi_1"] \\
\pi_2(\targetsp,\targetsp^1) \ar[r,"\p_\targetsp"]& \pi_1(\targetsp^1)
\end{tikzcd} \eeq
modulo a notion of homotopy between homomorphisms~\cite{APandJP}. Homotopy classes of homomorphisms between fundamental crossed modules can in fact be expressed in terms of cohomology groups~\cite{whitehead1949b}
\begin{framed}  
\begin{prop} \label{prop:2d}
The equivalence classes of homomorphisms between the fundamental crossed modules of $\basesp$ and $\targetsp$ can be expressed as
\beq [\Pi_{\leq 2}(\basesp),\Pi_{\leq 2}(\targetsp)]_0 = \bigcup_{\phi_1}H^2_{\phi_1}(\basesp,\pi_2(\targetsp)), \label{eqn.2dcoh} \eeq
where the union is over homomorphisms of fundamental groups $\phi_1\in[\pi_1(\basesp),\pi_1(\targetsp)]$. The cohomology groups have local coefficients, i.e. $\pi_2(\targetsp)$ should be viewed as a module over $\pi_1(\basesp)$ via $\phi_1$ and the action of $\pi_1(\targetsp)$ on $\pi_2(\targetsp)$.
\end{prop}
\end{framed}

\subsection{Examples}
\subsubsection{Warm up: Defects and textures in Heisenberg magnets}
Heisenberg magnets are characterized by a target space of a 2-sphere $\targetsp = SO(3)/SO(2) \simeq \bS^2$, corresponding to the spontaneous breaking of $SO(3)$ rotation symmetry to $SO(2)$ axial symmetry. Since $\pi_1(\bS^2)$ is trivial, there are no codimension two defects (nor textures on $\bS^1$). Meanwhile, codimension three defects (e.g. point defect in $\bR^3$) have the same classification as textures on $\bS^2$ since $\bR^d$ with $\bR^{d-3}$ removed can be continuously deformed to $\bS^2$. The topological sectors are
\beq \left[\bS^2,\bS^2\right] =\pi_2(\bS^2) \cong \bZ. \eeq 
These textures are known as magnetic skyrmions and the integer classifying them is known as the skyrmion charge. We can reproduce this using proposition \ref{prop:main}. The fundamental crossed module $\Pi_{\leq 2}(\bS^2)$ is $\bZ\to 0$, so crossed module homomorphisms of $\Pi_{\leq 2}(\bS^2)$ to itself are simply given by group homomorphisms $[\bZ,\bZ]\cong\bZ$.
\begin{figure}[!htb]
    \centering
        \includegraphics[width=4.1cm]{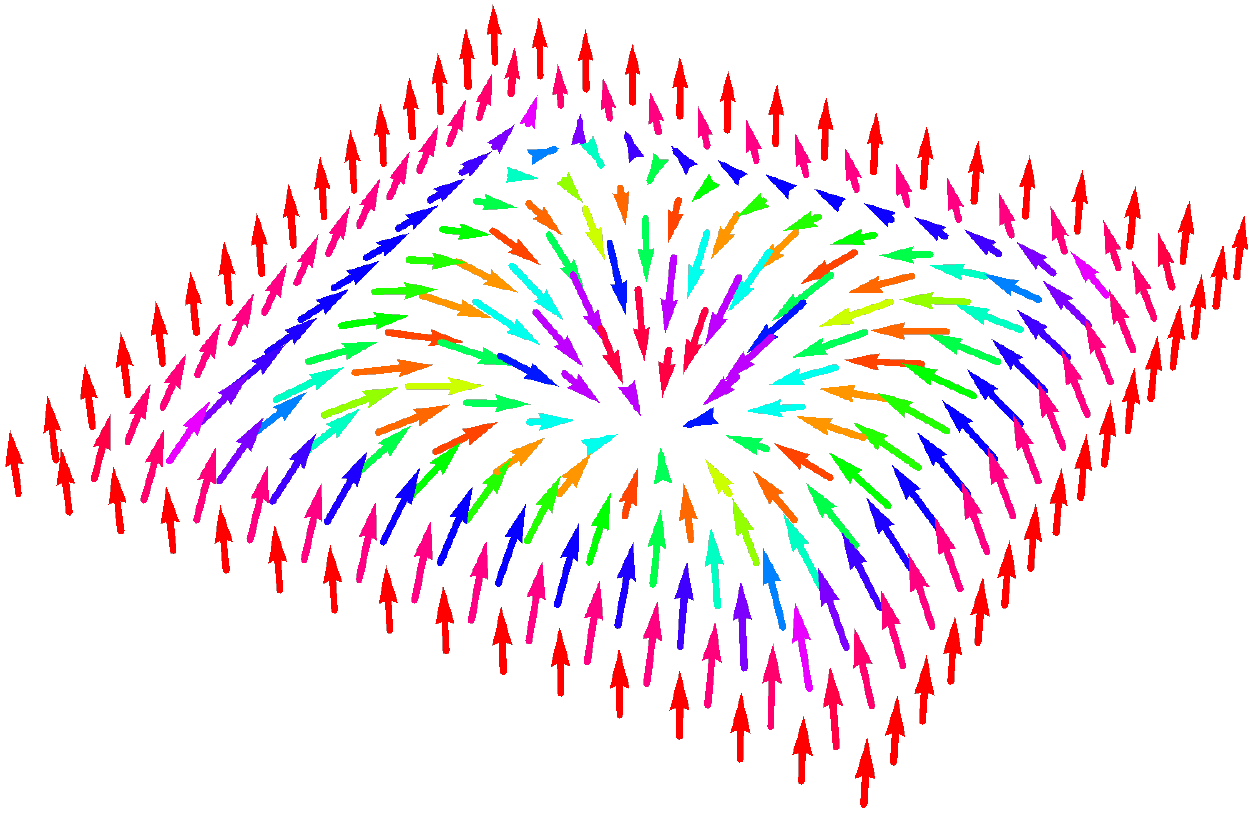} \includegraphics[width=4.1cm]{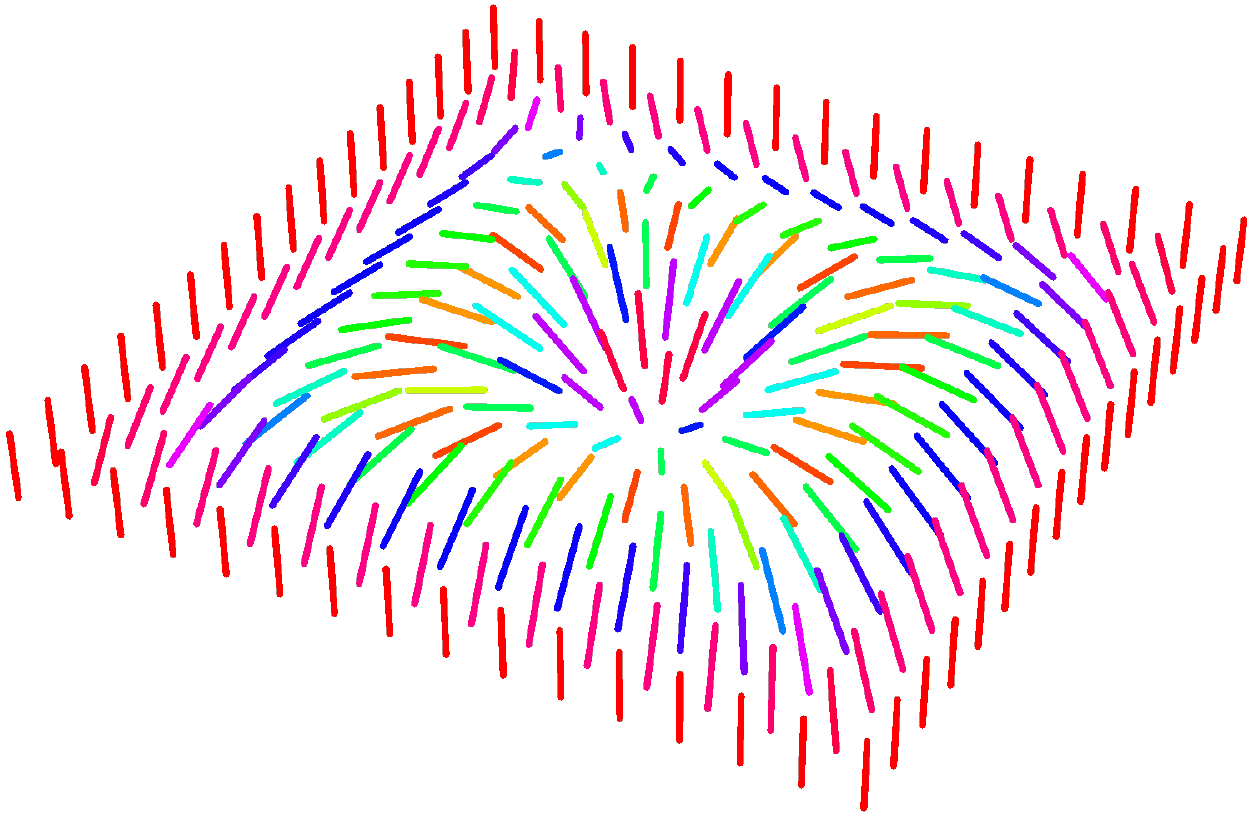}
    \caption{(Color online) Magnetic (left) and nematic (right) ``hedgehog" skyrmions of unit charge. By identifying the boundary square to a single point or by imposing periodic boundary conditions, these can be thought of as non-trivial textures on a sphere or torus respectively.}
    \label{fig:skyrmion}
\end{figure}

Next, consider a 2-torus $\basesp=\bT^2$. Homotopy classes $[\bT^2,\bS^2]$ are no longer given by a homotopy group, so we have to use proposition \ref{prop:main} or \ref{prop:2d}. The fundamental crossed module $\Pi_{\leq 2}(\bT^2)$ is the inclusion $G'\hookrightarrow G$, where $G$ is the free group on two generators, and $G'$ is its derived subgroup.\footnote{The derived subgroup $G'$ of $G$ is the subgroup generated by commutators of $G$.} Crossed module homomorphisms from $\Pi_{\leq 2}(\bT^2)$ to $\Pi_{\leq 2}(\bS^2)$ are determined by the image of any one commutator, so
\beq \left[\bT^2,\bS^2\right] \cong \bZ. \eeq
This coincides with the cohomology group $H^2(\bT^2,\bZ)\cong\bZ$ (c.f. proposition \ref{prop:2d}). These are also essentially skyrmions. Magnetic skyrmions on a 2-sphere and 2-torus are visualized on the left in fig.(\ref{fig:skyrmion}) by representing the order parameters as an arrow of fixed length.

\subsubsection{Torus textures in nematic liquid crystals}
Nematic liquid crystals are characterized by target space $\targetsp=SO(3)/O(2)=\RP^2$, corresponding to spontaneous symmetry breakdown of $SO(3)$ rotation symmetry to $O(2)=D_\infty$ symmetry. This model can host a rich variety of textures and defects. The results on two-dimensional nematic textures listed here have been studied elsewhere.~\cite{NamaticsAlexander,NematicsMachon20160265}

The first two homotopy groups of $\RP^2$ are $\pi_1(\RP^2) \cong \ztwo$ and $\pi_2(\RP^2) \cong \bZ$, and the non-trivial element of $\pi_1$ acts on $\pi_2$ by negating each integer $n\mapsto-n$. Therefore, codimension two defects are classified by $\left[\bS^1, \RP^2\right] \cong \ztwo$, while codimension three defects are classified by $\left[\bS^2,\RP^2 \right] \cong \left[\bS^2,\RP^2\right]_0/(n\mapsto -n) \cong \bN$. Unlike magnetic skyrmions, nematic skyrmions can only be identified up to the absolute value of the skyrmion charge $\abs{n}$. A nematic skyrmion on the 2-sphere is visualized on the right in fig.(\ref{fig:skyrmion}) by representing the order parameters as a rod-like director of fixed length.

For textures on a 2-torus $\bT^2$, homotopy groups are no longer sufficient. According to proposition \ref{prop:2d},
\beq
\left[\bT^2, \RP^2\right]_0 \cong \bigcup_{\phi_1}H^2_{\phi_1}(\bT^2,\bZ).
\eeq
Homomorphisms from $\pi_1(\bT^2)=\bZ^2$ to $\pi_1(\RP^2)=\bZ_2$ are characterized by the image $(x,y)$ of the two generators of $\bZ^2$. There are four choices, $(x,y) = (0,0),(0,1),(1,0),(1,1)$, where we write $\ztwo=\{0,1\}$ with additive notation. The cohomology groups twisted by these homomorphisms are
\bea
H^2_{(0,0)}(\bT^2,\bZ) &=& \bZ,\phantom{_2}~~~ H^2_{(1,0)}(\bT^2,\bZ)=\bZ_2, \nonumber \\
H^2_{(0,1)}(\bT^2,\bZ)&=&\bZ_2,~~~ H^2_{(1,1)}(\bT^2,\bZ)=\bZ_2. 
\eea
yielding
\beq \left[\bT^2,\RP^2\right]_0 = \bZ \cup \bZ_2 \cup \bZ_2 \cup \bZ_2. \eeq
\begin{figure*}[!htb]
    \centering 
    \begin{tabular}{cccc}
        \includegraphics[width=4cm]{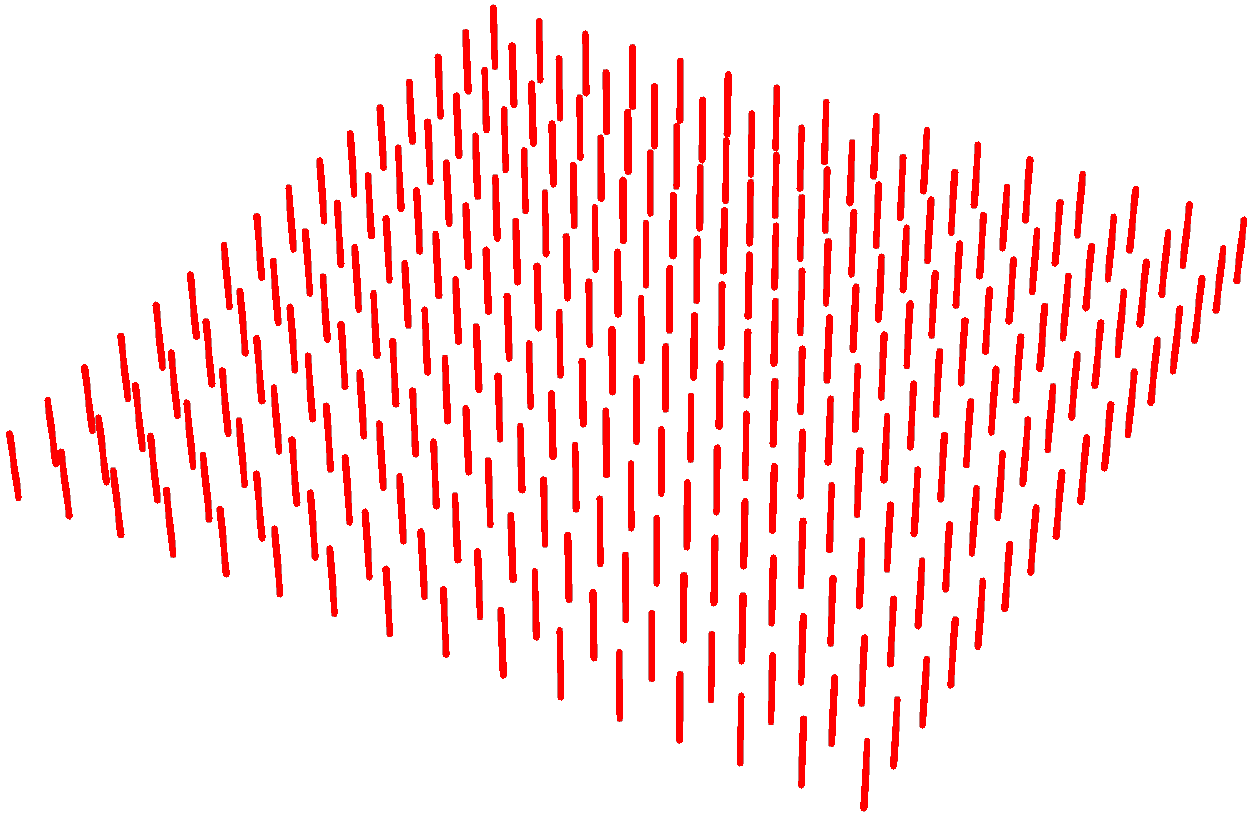} &     \includegraphics[width=4cm]{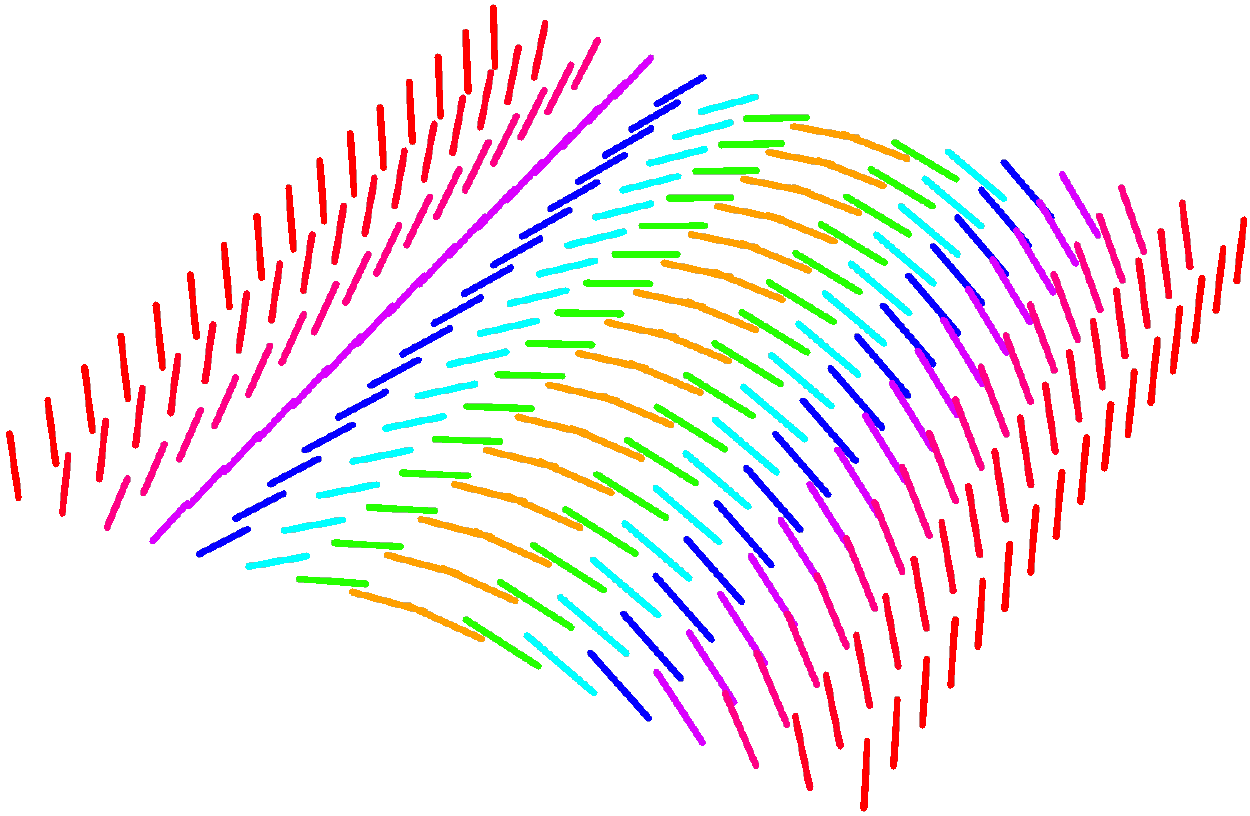} &
        \includegraphics[width=4cm]{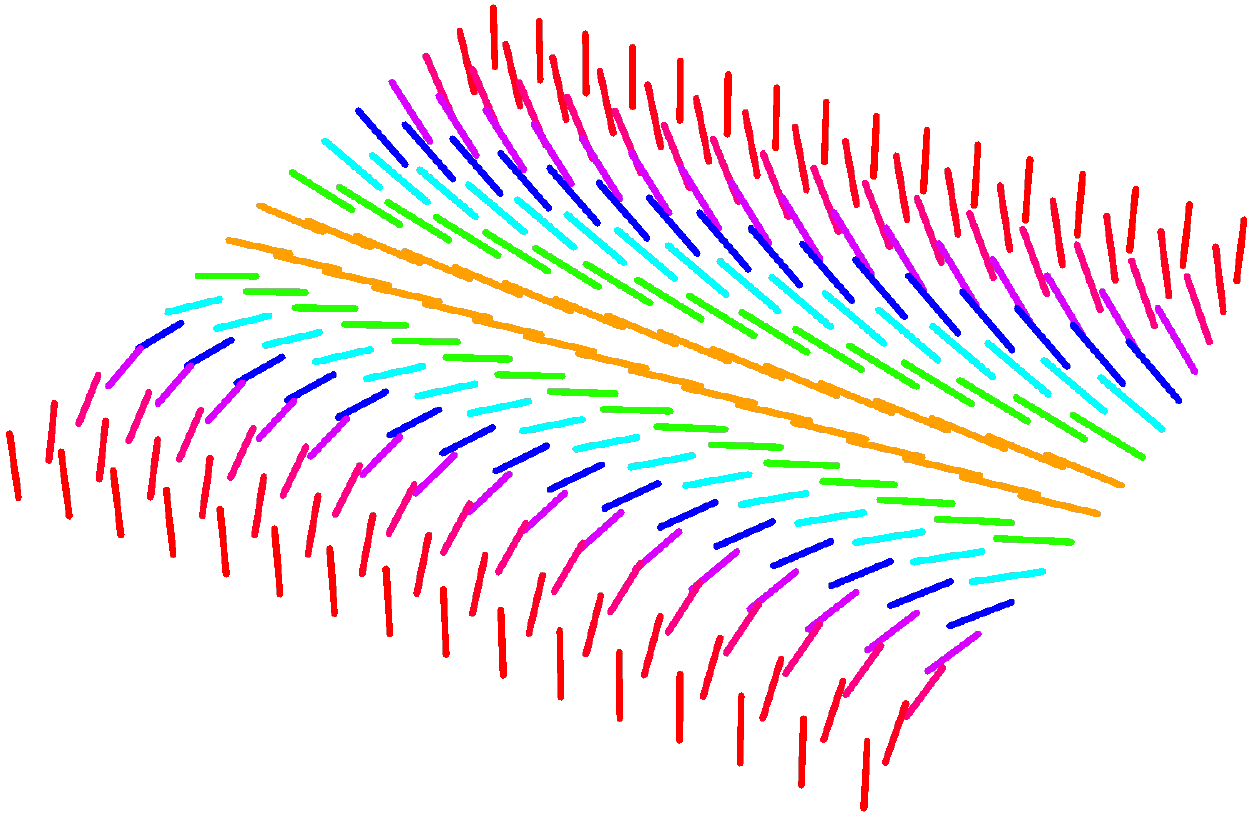} &
        \includegraphics[width=4cm]{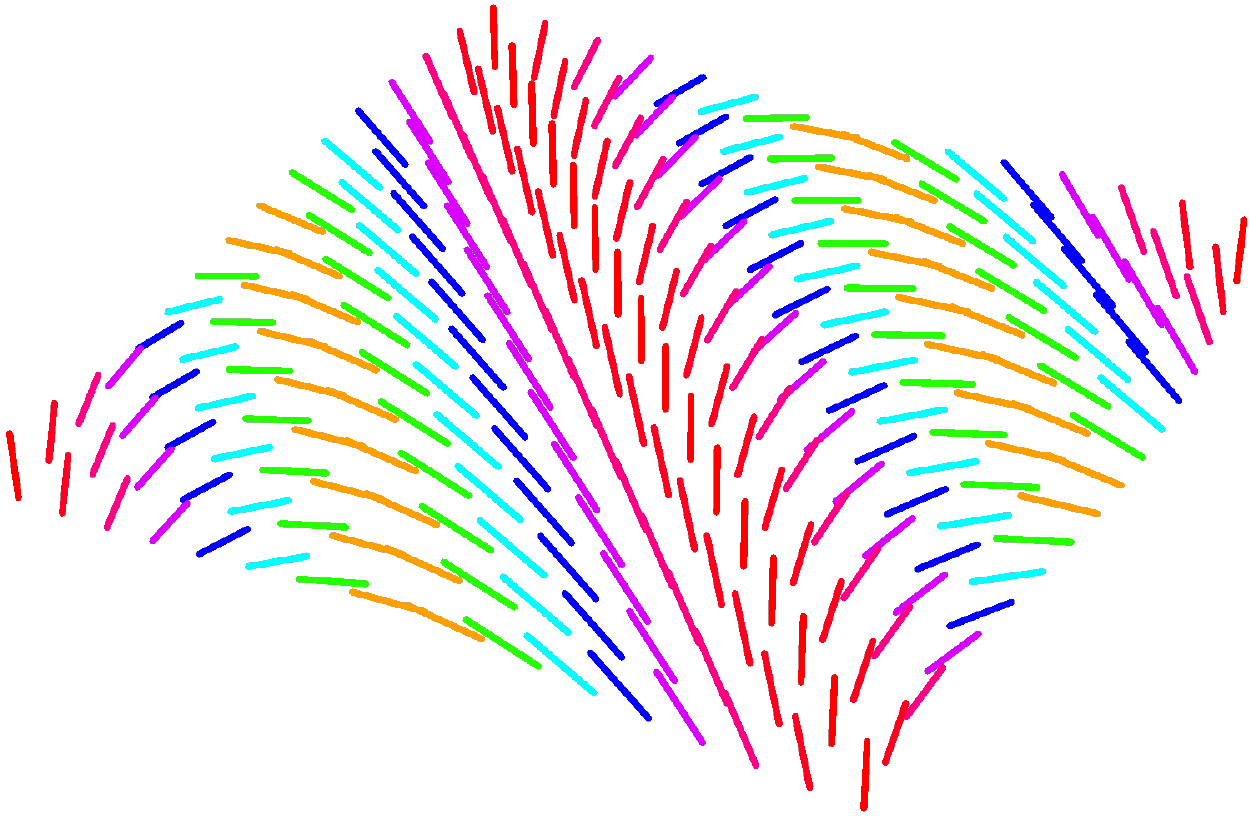} \\
        \includegraphics[width=4cm]{nematic_skyrmion.pdf} &     \includegraphics[width=4cm]{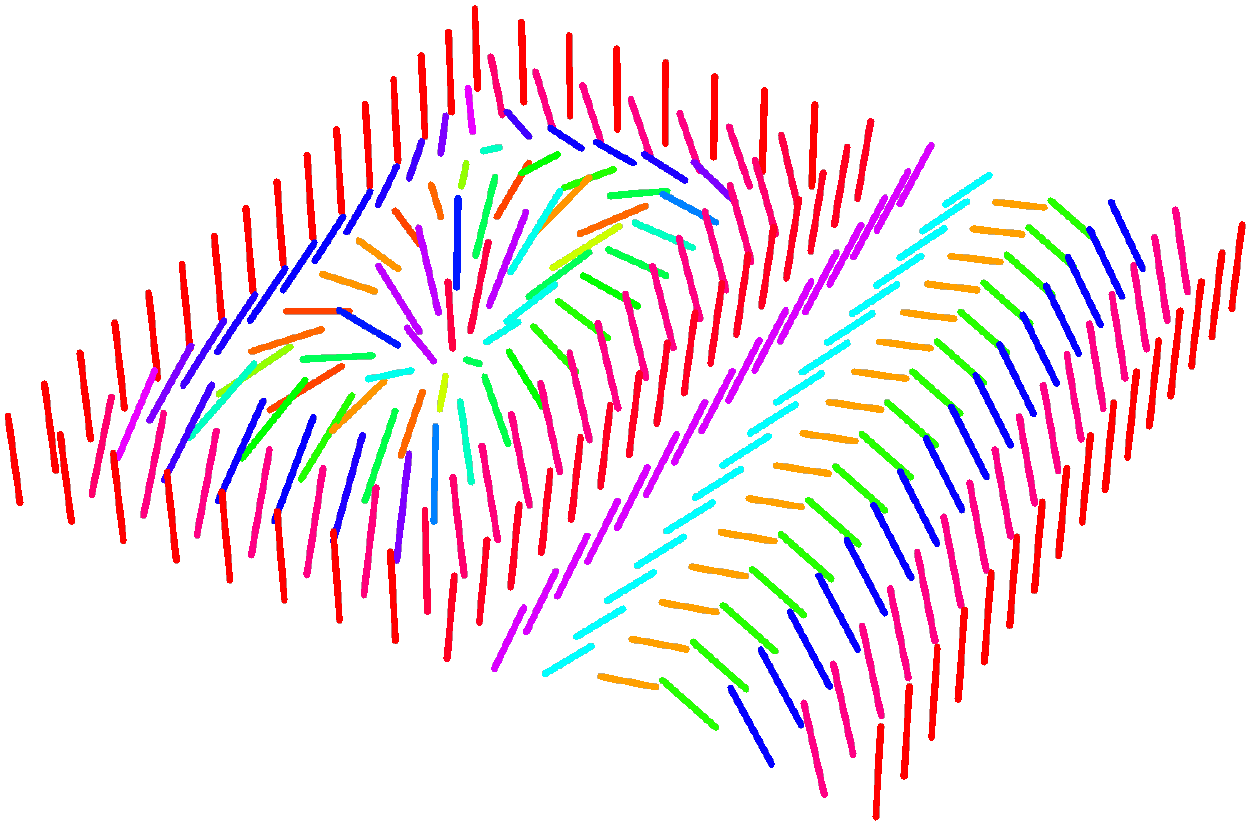} &
        \includegraphics[width=4cm]{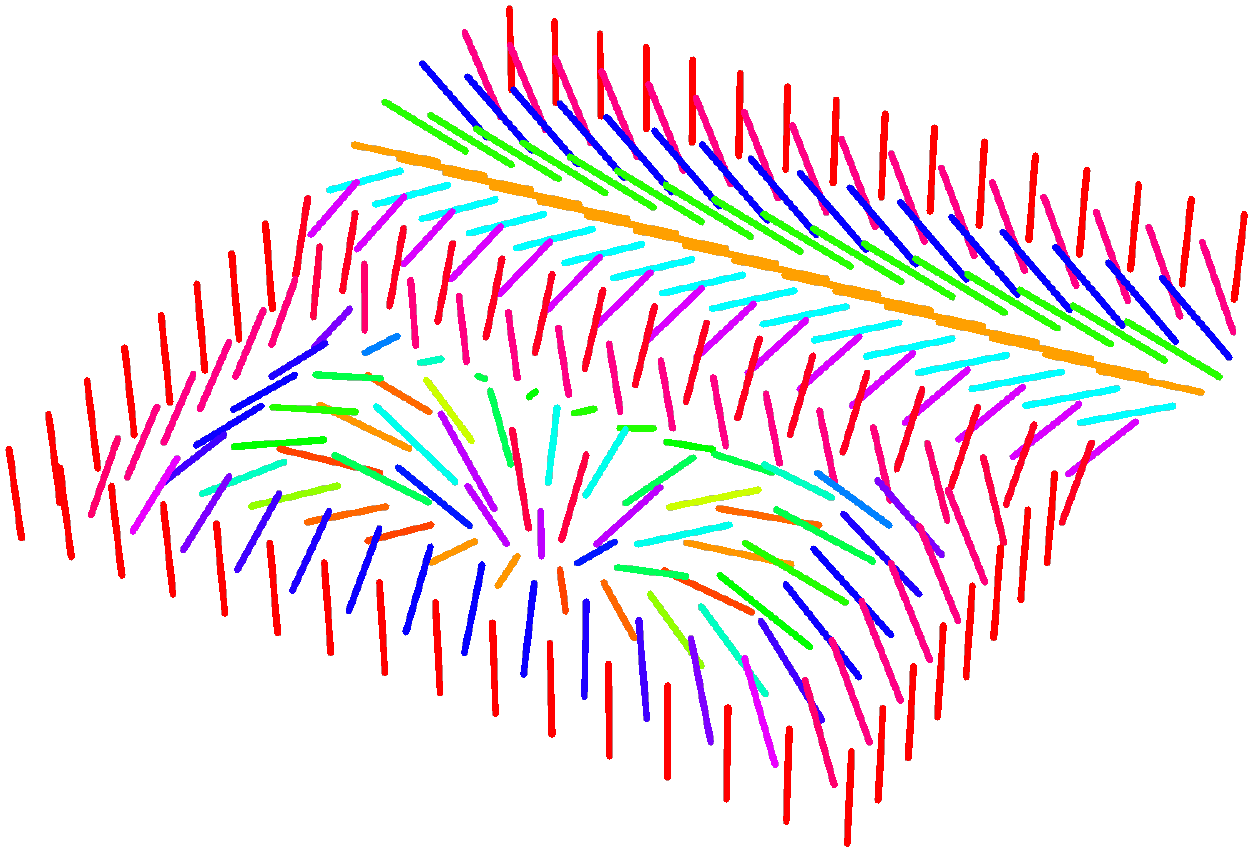} &
        \includegraphics[width=4cm]{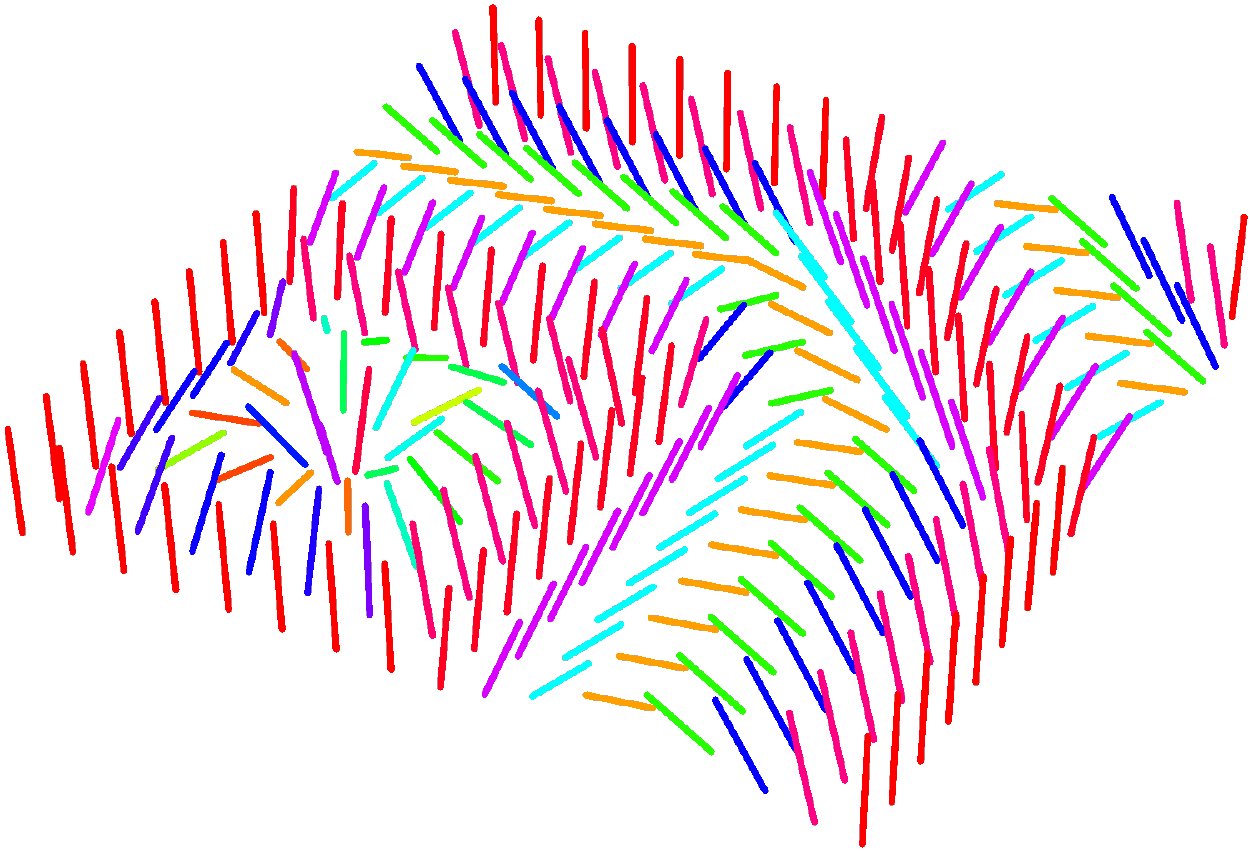}            
    \end{tabular}
    \caption{(Color online) Textures corresponding to the four $\phi_1:\pi_1(\bT^2) \to \pi_1(\RP^2)$}
    \label{fig:nematic sectors}
\end{figure*}
These textures can be visualized as follows. First, $H^2_{(0,0)}(\bT^2,\bZ) = \bZ$ are exactly the nematic skyrmions of integer charge discussed above. Next, one of the $\ztwo$ elements of each of $H^2_{(0,1)}(\bT^2,\bZ)$, $H^2_{(1,0)}(\bT^2,\bZ)$ and $H^2_{(1,1)}(\bT^2,\bZ)$ correspond to the last three figures in the top row of fig.(\ref{fig:nematic sectors}), where the nematic order parameter is represented by a rod-like director. The order parameters wind as they go around the incontractible cycles of the torus, in a manner determined by $\phi_1$. If a skyrmion is introduced into one of these configurations, transporting it around a cycle where the texture winds non-trivially induces an action of $\pi_1$ on $\pi_2$, negating the skyrmion charge. Thus, any skyrmion with even charge can be annihilated by splitting it into two equally charged skyrmions and transporting one of them around the cycle. This means that the three textures can each be modified by introducing a single skyrmion (see the last three figures on the bottom row of fig.(\ref{fig:nematic sectors})), but introducing two yields a configuration which can be smoothly deformed to the original. This explains why there are only two distinct textures corresponding to each of $H^2_{(0,1)}(\bT^2,\bZ)$, $H^2_{(1,0)}(\bT^2,\bZ)$ and $H^2_{(1,1)}(\bT^2,\bZ)$.

Finally, we relax the basepoint condition by considering the effect of $\pi_1$. From the analysis above, we can see that this amounts to identifying skyrmion charges $n$ and $-n$, which only affects the sector with trivial $\phi_1$, yielding
\beq \left[\bT^2,\RP^2\right] = \bN \cup \bZ_2 \cup \bZ_2 \cup \bZ_2. \eeq


\subsubsection{Loop and knot defects in nematic liquid crystals}
\begin{figure}
    \centering
    \includegraphics[width=0.3\textwidth]{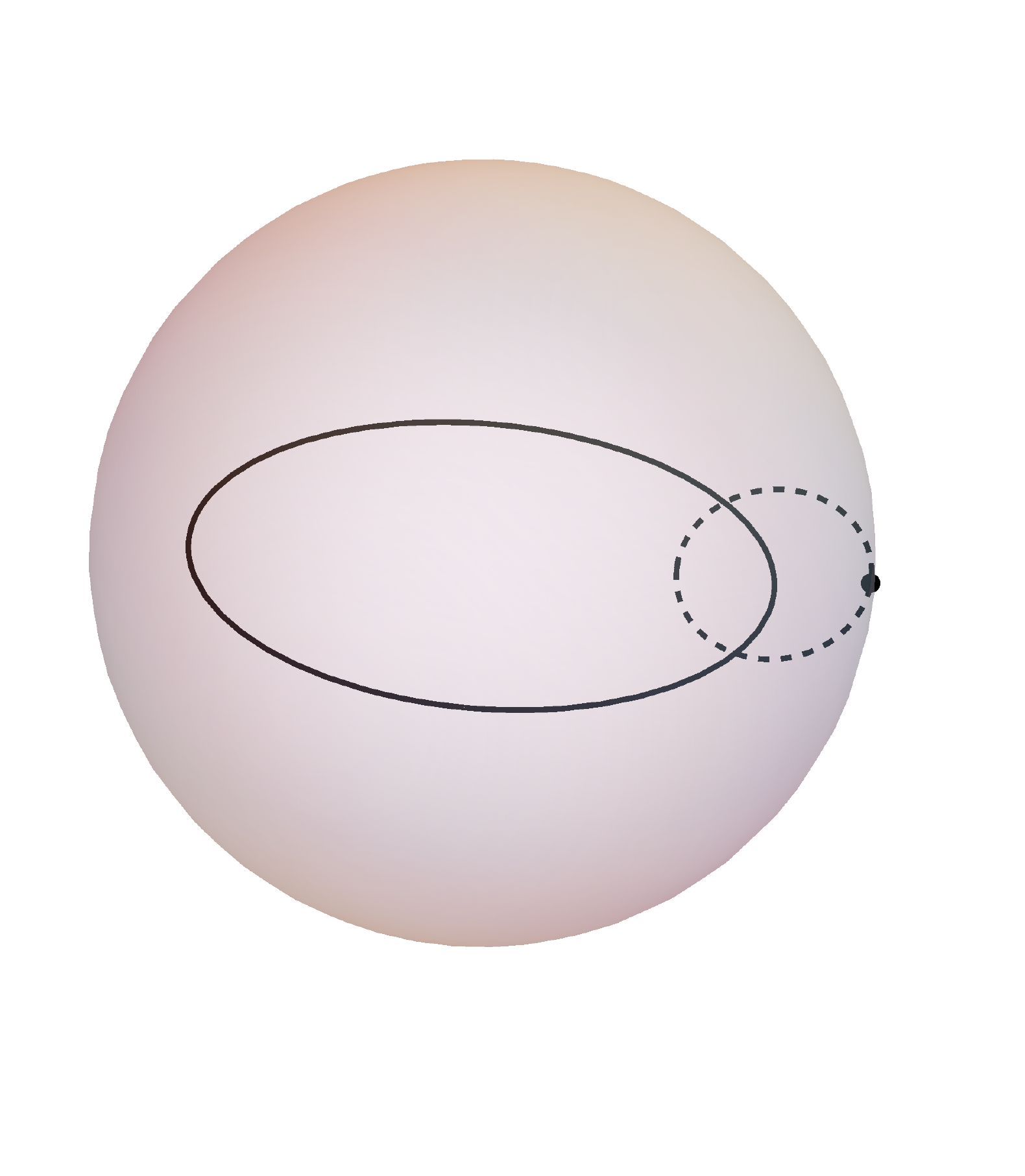}
    \caption{The complement of an unknotted loop defect retracts onto $\bS^1\vee\bS^2$.}
    \label{fig:loop}
\end{figure}

In this example, we consider defects supported on a loop in $\bR^3$, first in the unknotted case, and later allowing the loop to form non-trivial knots. When it is unknotted, the complement $\bR^3\setminus\bS^1$ retracts onto $\basesp=\bS^1 \vee \bS^2$, which denotes $\bS^1$ and $\bS^2$ with a common point (see fig.(\ref{fig:loop}). The circle $\bS^1$ links the loop defect, and the sphere $\bS^2$ encircles the entire loop. A straightforward computation of the crossed module homomorphisms yields~\cite{APandJP}, for generic target space $\targetsp$,
\beq
\left[\bS^1 \vee \bS^2, \targetsp \right]_0 \cong \pi_1(\targetsp)\times\pi_2(\targetsp),
\eeq
reproducing the result obtained previously by Nakanishi et al.~\cite{Nakanishi1988}. Specifying to nematics, the result is
\beq
\left[\bS^1 \vee \bS^2, \RP^2 \right]_0 \cong \bZ \cup \bZ.
\eeq
Finally, relaxing the based condition by incorporating the $n\mapsto -n$ action of $\pi_1(\RP^2)$ yields
 \beq
\left[\bS^1 \vee\bS^2, \RP^2 \right] \cong \bN \cup \bN.
\eeq
This can be thought of as the skyrmion charge of the loop defect when viewed as a point defect at large distance, with the two copies corresponding to whether or not the nematic winds non-trivially around a circle linking the loop defect.

Next, consider knotted defects. We consider them in $\bS^3$ rather than $\bR^3$, which corresponds to demanding that the configuration asymptotes to a fixed point in $\RP^2$ at infinity. Should we choose to view them in $\bR^3$, there will be an extra factor of $\bN$ corresponding to the skyrmion charge of the entire defect when viewed as a point defect at large distance, or, equivalently, as the ``skyrmion charge at infinity''. For simplicity, we restrict to knots that can be inscribed on a torus embedded in $\bS^3$. They are labeled by two coprime integers $(p,q)$ which are the number of times the knot winds along each cycle of the torus before reconnecting with itself. The complement $M_{p,q}$ of a $(p,q)$ torus knot in $\bS^3$ has fundamental group $\pi_1(M_{p,q}) \cong \innerproduct{a,b}{a^p = b^q}$ and $\pi_2=0$~\cite{lickorish1997}. For knotted defects in nematic liquid crystals $\targetsp=\RP^2$, the homomorphisms of fundamental groups $[\pi_1(M_{p,q}),\bZ_2]$ depends on whether $p$ and $q$ are even or odd. We analyze the three possible cases in turn.
\begin{enumerate}
    \item $p$ even and $q$ odd:  $\phi_1(a)$ can be assigned either element of $\ztwo$ and $\phi_1(b)$ can only be assigned the trivial element. According to proposition \ref{prop:2d},
    \begin{align}
    H^2_{(0,0)}(M_{p,q},\bZ) \cong 1,~~~&H^2_{(1,0)}(M_{p,q},\bZ) \cong \bZ_p \\
    \left[M_{p,q},\RP^2\right]_0 &\cong 1 \cup \bZ_p 
    \end{align}
    \item $p$ odd and $q$ even:  $\phi_1(b)$ can be assigned either element of $\ztwo$ and $\phi_1(a)$ can only be assigned the trivial element.  
    \begin{align}
        H^2_{(0,0)}(M_{p,q},\bZ) \cong 1,~~~&H^2_{(0,1)}(M_{p,q},\bZ) \cong \bZ_q \\
    \left[M_{p,q},\RP^2 \right]_0 &\cong 1 \cup \bZ_q 
    \end{align}
    \item $p$ and $q$ odd: $\phi_1(a)$ and $\phi_1(b)$ both have to be assigned to the same element of $\ztwo$.  
    \begin{align}
        H^2_{(0,0)}(M_{p,q},\bZ) \cong 1,~~~&H^2_{(1,1)}(M_{p,q},\bZ) \cong 1 \\
    \left[M_{p,q},\RP^2 \right]_0 &\cong 1 \cup 1
    \end{align}
\end{enumerate}
This agrees with previous results~\cite{NematicsMachon20160265} that the number of non-trivial based defect sectors in a nematic liquid crystal with knot defect is equal to the knot determinant.

To relax the based constraint, we need to understand the action of $\pi_1$ on these sectors. This is explained in~\cite{APandJP}. As an example, the $(p,q)=(2,3)$ torus knot, also known as the trefoil knot, has four based defect sectors, two of which are identified through an action of $\pi_1$, leaving three distinct defect sectors.

\section{$\basesp$ with homotopy dimension $\ge$3}
Whitehead's result~\cite{whitehead1949b,ellis1988} was actually more general than what we stated in proposition~\ref{prop:2d}. It applied to arbitrary dimensions, as long as the target space satisfied extra constraints:

\begin{framed}
\begin{prop}\label{prop:Whitehead}
Let $\basesp$ be of homotopy dimension $d$, and $\targetsp$ be such that $\pi_i(\targetsp)=0$ for $i=2,3,\ldots,d-1$. Then
\beq  \left[\Pi_{\leq d}(\basesp),\Pi_{\leq d}(\targetsp)\right]_0 \cong \bigcup_{\phi_1} H_{\phi_1}^d(\basesp,\pi_d(\targetsp)), \eeq
\noindent where $\phi_1\in[\pi_1(\basesp),\pi_1(\targetsp)]$ are homomorphisms of fundamental groups, and the cohomology groups have local coefficients via $\phi_1$ and the action of $\pi_1(\targetsp)$.
\end{prop}
\end{framed}

When the constraint that the intermediate homotopy groups of $\targetsp$ are trivial is not satisfied, the situation becomes much more complicated and one has to count the equivalence classes of crossed square homomorphisms
\beq \begin{tikzcd}[row sep=scriptsize,column sep=tiny]
& \pi_3(M;M^2_+,M^2_-) \arrow[dl] \arrow[rr] \arrow[dd,"\phi_3",near start] & & \pi_2(M^2_+,W) \arrow[dl] \arrow[dd,"\phi_2^+"] \\
\pi_2(M^2_-,W) \arrow[rr, crossing over] \arrow[dd,"\phi_2^-"] & & \pi_1(W) \\
& \pi_3(X;X^2_+,X^2_-) \arrow[dl] \arrow[rr] & & \pi_2(X^2_+,Y) \arrow[dl] \\
\pi_2(X^2_-,Y) \arrow[rr] & & \pi_1(Y) \arrow[from=uu, crossing over,"\phi_1",near start]\\
\end{tikzcd} \eeq
as in proposition \ref{prop:main}. The notation in the above commuting diagram is explained in~\cite{APandJP}. In the remainder of this section, we discuss examples of textures in three dimensions, first when the constraint in proposition \ref{prop:Whitehead} is satisfied, and later when it is not.

\subsection{Examples \'{a} la Whitehead}
\subsubsection{Target space $\bS^3$}
Textures on $\basesp=\bS^3$ or $\bT^3$ are three dimensional skyrmions, classified by integer charge.
\bea
\left[\bS^3,\bS^3\right] &\cong& \pi_3(\bS^3) \cong \bZ \\
\left[\bT^3,\bS^3\right] &\cong& H^3(\bT^3,\pi_3(\bS^3)) \cong \bZ 
\eea
\subsubsection{Target space $\RP^3$}
The order parameter of the spiral phase of quantum antiferromagnets takes values in $SO(3)$~\cite{PhysRevB.84.104430}, whose underlying manifold is $\RP^3$. Its first three homotopy groups are $\pi_1(\RP^3) \cong \ztwo$, $\pi_2(\RP^3)\cong 0$ and $\pi_3(\RP^3) \cong \bZ$. $\pi_1$ acts trivially on $\pi_3$.

Textures on $\bS^1$ and codimension 2 defects in $\bR^d$ are classified by $\pi_1(\RP^3)\cong \ztwo$. There are no non-trivial textures on $\bS^2$ nor codimension 3 defects. On the 2-torus $\bT^2$, since
\beq \left[\bT^2,\RP^3\right] \cong \bigcup_{\phi_1}H^2_{\phi_1}(\bT^2,\pi_2(\RP^3)) \cong \bigcup_{\phi_1}1, \eeq
there are precisely four textures given by the four homomorphisms $\phi_1\in[\pi_1(\bT^2),\pi_1(\RP^3)]=[\bZ^2,\bZ_2]$. This is identical to the possible choices of $\pi_1(\bT^2)\to\pi_1(\RP^2)$ in the case of the nematic liquid crystal, but in this case there is only one texture for each choice of $\phi_1$, since, intuitively, there is no cycle for a surface to wrap.

Moving on to three dimensional textures, note that since $\pi_2(\RP^3)=0$, it satisfies the conditions of proposition \ref{prop:Whitehead}. Textures on $\bS^3$, or codimension 4 defects, are classified by
\beq \left[\bS^3,\RP^3\right] \cong \pi_3(\RP^3) \cong \bZ. \eeq
These textures correspond to higher dimensional skyrmions. According to proposition \ref{prop:Whitehead}, textures on $\bT^3$ are classified by
\beq \left[\bT^3,\RP^3\right]_0 \cong \bigcup_{\phi_1}H^3_{\phi_1}(\bT^3,\pi_3(\RP^3)). \eeq
The eight homomorphisms $\phi_1: \pi_1(\bT^3) \to \pi_1(\RP^3)$ are characterized by the images of each of the three generators of $\pi_1(\bT^3)\cong\bZ^3$. Since $\pi_1$ acts trivially, the cohomology groups are taken with constant coefficients,
\beq H^3_{\phi_1}(\bT^3,\bZ) = \bZ. \eeq
The based sectors can be visualized as textures that wind around the three non-contractible cycles of $\bT^3$ as determined by $\phi_1$, and from there, the rest of the configurations can be obtained by introducing skyrmions, just as was done for the nematic liquid crystal $\RP^2$. In this case, taking a skyrmion around a winding cycle does not change the skyrmion charge, even when $\phi_1$ is non-trivial. Therefore, relaxing the based constraint does not identify any of the configurations. The three dimensional textures of the spiral phase of antiferromagnets, with periodic boundary conditions, are given by
\beq
\left[\bT^3,\RP^3\right] = \bigcup^8 \bZ = (\bZ_2)^3\times\bZ.
\eeq

\subsection{Magnetic textures in 3D}
We saw earlier that Heisenberg magnets give rise to non-trivial textures and defects when $\basesp$ is two dimensional. When $\basesp$ is three dimensional, even more interesting textures and defects can be observed. Since $\pi_2(\bS^2)$ is nonzero, topological sectors of Heisenberg magnets fall outside the scope of prop.~\ref{prop:Whitehead}. We have to count equivalence classes of crossed square homomorphisms, as in prop.~\ref{prop:main}. The fundamental crossed $2$-cube, or square, denoted $\Pi_{\le 3}(\basesp)$, contains all the homotopy information of a three dimensional CW complex $\basesp$. We do not define fundamental crossed squares $\Pi_{\le 3}(\basesp)$ in this letter, but instead present two examples below as an invitation to~\cite{APandJP}. These results were obtained earlier~\cite{Pontrjagin1941} in the mathematics literature using a different method.

Textures on $\bS^3$ are classified by
\beq
\left[\bS^3,\bS^2 \right] \cong \pi_3(\bS^2) \cong \bZ. 
\eeq
These textures are called Hopf solitons or simply hopfions in the literature. They can be interpreted as pair production of skyrmions which braid around each other a number of times before annihilation~\cite{wilczek1983}.

Our first example is $\basesp=\bS^1 \times \bS^2$, which can be thought of as a thermal theory of a point defect in three spatial dimensions. Computing the topological sectors following eq.(\ref{eq:d dim based sectors 2}) yields the result~\cite{APandJP}
\beq \left[ \bS^1 \times \bS^2, \bS^2\right] \cong \bigcup_{q \in \bZ}  \bZ_{|2q|} \eeq
with $\bZ_0$ interpreted as $\bZ$. The integer $q$ labels the skyrmion charge on any $\bS^2$ cross-section. The other textures can be thought of as being produced by the introduction of a hopfion to this background. For $q=0$, the case is identical to textures on $\bS^3$ and introducing any number of hopfions leads to distinct textures, while for $q \neq 0$, the introduction of $\abs{2q}$ hopfions leads to a texture which can be deformed back to the original, leaving only $\abs{2q}$ distinct textures.

The second example is $\basesp=\bT^3$, which can be thought of as a three dimensional texture with periodic boundary conditions. The result is
\beq
\left[\bT^3,\bS^2 \right] \cong \bigcup_{(q_1,q_2,q_3) \in \bZ^3} \bZ_{2\gcd(\abs{q_1},\abs{q_2},\abs{q_3})}.
\eeq
$(q_1,q_2,q_3)$ can be thought of as the skyrmion charge on each of the three orthogonal $\bT^2$ cross sections. Additional textures are produced by introducing hopfions to these backgrounds. Once again, when $q_1=q_2=q_3=0$, the case is identical to textures on $\bS^3$ and introducing any number of hopfions leads to distinct textures. When $(q_1,q_2,q_3)\neq(0,0,0)$, some textures can be smoothly deformed into each other in an interesting way, leaving $2\gcd(\abs{q_1},\abs{q_2},\abs{q_3})$ distinct classes.

\section{Relation to higher categorical groups in physics}
Higher categorical generalizations of groups have recently found several applications in physics -- amongst both the high energy and condensed matter communities. Higher groups have been studied as generalizations of global~\cite{GeneralizedGlobalSymmetriesGaiotto2015} and gauge~\cite{HigherSymmetreisGappedPhasesKapustinThorngren2017} symmetries. Just like ordinary symmetries, their presence can lead to phase transitions both with spontaneous symmetry breaking~\cite{Lake:2018dqm,Gukov:2013zka} and without~\cite{HigherSymmetreisGappedPhasesKapustinThorngren2017}. They can also be anomalous, which imposes powerful constraints on the IR behaviors of theories and hence be used to test conjectured dualities~\cite{Benini:2017dus,2GroupAnomaliesBenini:2018reh,Cordova:2017kue}. The Green-Schwarz mechanism can also be reinterpreted as non-trivial 2-group gauge symmetry~\cite{GreenSchwartzCordova:2018cvg}. The classification of symmetry-enriched topological (SET) phases in 2+1 dimensions~\cite{Barkeshli:2014cna} can be obtained by coupling ordinary global symmetries to the automorphism 2-group of the modular tensor category data characterizing the topological order of the system.~\cite{2GroupAnomaliesBenini:2018reh}. 

Fundamental crossed $(d-1)$-cubes are models for a certain type of higher categorical groups known as $(d-1)$-cat groups~\cite{SpacesWithFinitelyLODAY1982179}. In $d=2$, crossed $1$-cubes, or crossed modules, $\Pi_{\le 2}$ are equivalent to a flavor of 2-groups known as strict 2-groups~\cite{Baez2003,HigherSymmetreisGappedPhasesKapustinThorngren2017}. In $d=3$, it is unclear to us how $2$-cat groups are related to other variants of higher categorical groups, and will be the subject of future study.

In this work, we have demonstrated another instance where higher categorical generalizations of groups provide a natural language to understand physical phenomena. One difference from the other work described above, is that the higher categorical groups do not describe symmetries of the theory in this case.

\section{Conclusions and outlook}
In this work, we have studied a framework based on higher categorical generalization of groups to classify topological textures and defects. We also work out several examples, some known and some new, to make the exposition clear.  

There are several avenues where we anticipate our results might be useful. Topological theta terms probe topological sectors of a theory and lead to interference between them~\cite{ABANOVthetaterms,Abanov2017}. It would be interesting to study them using our results. Since topological sectors are also useful in studying anomalies~\cite{Intrinsic_emergent_anomalies}, deconfined criticality~\cite{deconfined_criticality} and in the classification of gapped phases of matter~\cite{surface_SPT,Statistical_Witten}, we hope our results can aid in the study of these interconnected problems.

\section{Acknowledgements}
We are grateful to Martin Ro\v{c}ek, Dennis Sullivan, Sasha Abanov, Ying Hong Tham, Tobias Shin, Sasha Kirillov and Onkar Parrikar for several helpful discussions. AP is grateful to the Kavli Institute for Theoretical Physics for their hospitality during the writing of this manuscript. JPA is supported in part by NSF grant PHY1620628. AP was supported in part by NSF 
grants PHY1333903, PHY1314748, and PHY1620252 and is currently supported by the Simons Foundation via the ICTS-Simons postdoctoral fellowship.

\bibliography{references}{}

\appendix

\end{document}